\documentclass[10pt]{iopart}
\usepackage{graphicx}  
\begin{document}

\title[Exotic hadrons from dynamical clustering of quarks]{Exotic hadrons from dynamical clustering of quarks in ultrarelativistic heavy ion collisions}

\author{Stefan Scherer\dag
\footnote[3]{scherer@th.physik.uni-frankfurt.de}
}

\address{\dag\ Institut f\"ur Theoretische Physik, 
Johann Wolfgang Goethe-Universit\"at,
Robert-Mayer-Str.\ 8-10,                
D-60054 Frankfurt am Main, Germany}

\begin{abstract}
Results from a model study on the formation of exotic quark clusters 
at the hadronization stage of a heavy ion collision are presented. 
The dynamical quark molecular dynamics (qMD) model which is used is sketched, 
and results for exotica made of up to six (anti-)quarks are shown. 
The second part focuses on pentaquarks. The rapidity distribution are shown, 
and the distribution of strangeness is found to yield an indicator of 
thermalization and homogenisation of the deconfined quark system. 
Relative $\Theta^+$ yields are found to be lower than thermal model
estimates.
\end{abstract}




\nosections
In the quark model of QCD, quarks are associated with fundamental 
representations of the colour-gauge group SU(3). The low energy regime of QCD 
contains only colour singlets such as mesons and baryons, which are made 
up from a quark and anti-quark with colour and anti-colour, or three quarks 
with three different colours, respectively. However, colour neutrality can 
be fulfilled whenever the net number of quarks is a multiple of three, thus
opening the possibility of so called exotic hadrons such as the tetraquark (or dimeson), 
the pentaquark, and so on. New interest in the study of exotic hadrons has been
created by recent reports about the discovery of pentaquark states $\Theta^+$ and $\Xi^{--}$ 
in pA, nA, and $\gamma$A collisions (\cite{experiment} -- see \cite{theory} for discussions of the 
theoretical understanding). 

Here, we will discuss the formation of clusters of up to six quarks or anti-quarks 
in the hadronization stage of a heavy ion collision, with a focus on pentaquarks. 
In such an event, two nuclei collide, creating hot and dense matter,
which can set free colour charges if the energy density is high enough. 
During the expansion of the hot system, colour charges will recombine
to form colour neutral hadrons which finally can be detected. The study is 
motivated by the idea that the hadronization from a hot quark soup is an environment 
favourable to the formation of exotica~\cite{Chen:2003tn}.

\begin{figure}[tbhp]
\centerline{\includegraphics[width=1.0\linewidth]{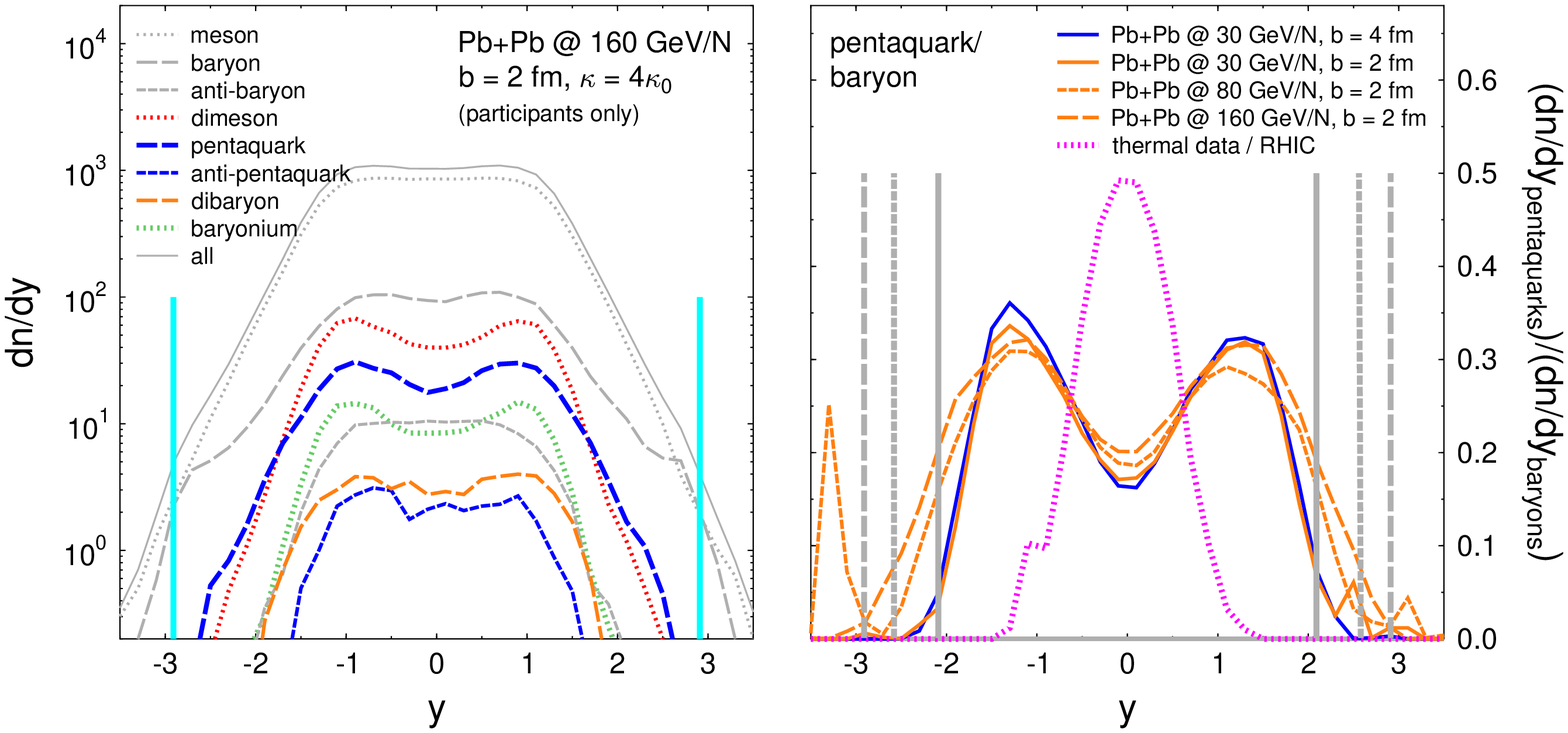}}
\caption{\label{FIG:1}
Rapidity distribution of different types of quark clusters at upper SPS energies (left figure).
Pentaquark clusters are represented by the thick doted line.
Rapidity distribution of pentaquark clusters, normalised to the total baryon distribution, for
different colliding systems (right figure).
}
\end{figure}

In order to describe the dynamics of the collision, the hadronic model UrQMD is
used to create an initial state of hot and dense matter. The particle content
after complete overlap of the nuclei is converted to quarks, and the quark molecular 
dynamics model qMD is used to describe the subsequent dynamics 
and recombination of the quark system~\cite{Hofmann:1999jx}. In the qMD model, 
quarks are described as classical particles with current masses subject to the Hamiltonian
\begin{equation} \label{EQ:1}
{\mathcal{H}} \,=\,
\sum_{i=1}^N\sqrt{\vec{p}_i^2+m_i^2} 
- \frac{1}{2}\sum_{i,j}C_{ij}^{\mathrm{c}} \left( \frac{3}{4}\frac{\alpha_s}{\left|\vec{r}_i-\vec{r}_j\right|}+\kappa\,\left|\vec{r}_i-\vec{r}_j\right|\right)\;.
\end{equation}
All quarks carry spin and isospin which is used in the subsequent mapping to hadrons, and 
colour charges $\vec{w}_{i}$ represented by three vectors in the root diagram, with
anti-quarks carrying the corresponding anti-colours. No colour exchange between quarks is assumed, 
leaving the colour of a particle unchanged throughout the complete time evolution of the system. 
The sign of the potential interaction between two particles depends on the scalar product of the 
root vectors,
\begin{equation} \label{EQ:2}
C_{ij}^{\mathrm{c}} \,=\, \vec{w}_{i}^{\sf T}\vec{w}_{j}\;,
\end{equation}
yielding repulsion for equal charges and attraction between charge and anti-charge.
The shape of the potential interaction is motivated by the Cornell potential, where in all 
calculations the Coulomb term is omitted, and the string constant $\kappa$ is treated as a 
parameter of the model. The time evolution yields an expanding system where quarks form 
colour neutral clusters. Once a cluster is separated in space from the remaining quarks
and its interaction with the system falls below a threshold, it is mapped to a hadron, 
taking into account the quantum numbers and energy-momentum four-vectors of the quarks 
in the cluster~\cite{Scherer:2001ap}.

\begin{figure}[tbhp]
\centerline{\includegraphics[width=1.0\linewidth]{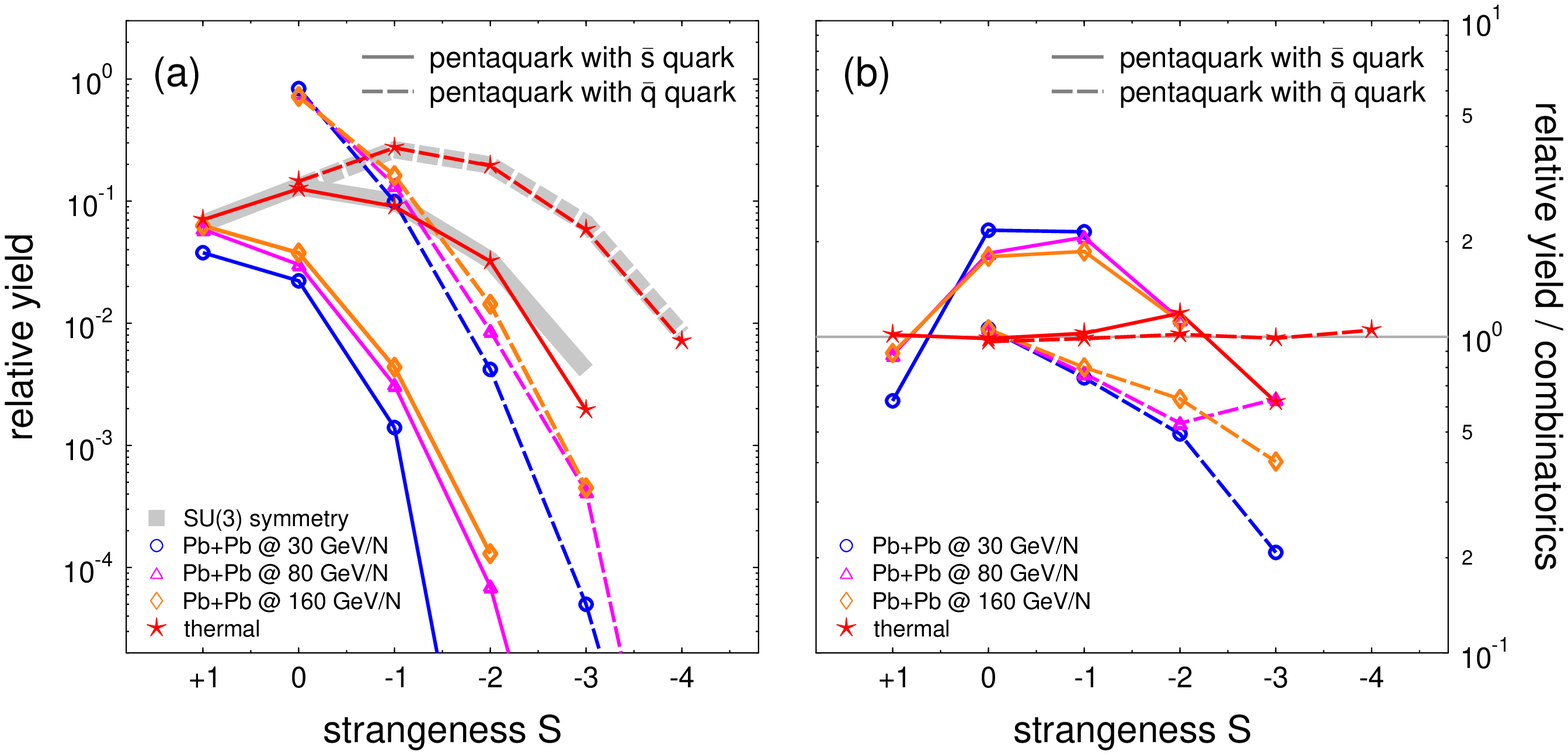}}
\caption{\label{FIG:3}
Distribution over strangeness of pentaquark cluster yield (a), and normalised to the 
combinatorial expectation value (b). Yields of $\Theta^+$ ($S=+1$ with a $\bar s$ quark) 
and $\Xi^{--}$ ($S=-2$ with a $\bar q$ quark) are only half of the combinatorial value
at lower SPS energies.
}
\end{figure}

We investigate the formation of colour neutral clusters with quark configurations of 
mesons ($q\bar q$), baryons ($qqq$), dimesons ($qq\bar q \bar q$), pentaquarks ($qqqq\bar q$), 
dibaryons ($qqqqqq$), baryonia ($qqq\bar q\bar q\bar q$), and the corresponding anti-particles.
We cannot make any statement about the actual existence and physical properties of exotic hadrons
from our extremely simplistic classical model, but we assume that any physical exotic hadron formed
by quark coalescence at hadronization should emerge from a colour neutral cluster as described 
in the qMD model. The dynamical evolution of systems at upper SPS energies then yields the rapidity
distribution of clusters as shown on the left part of figure \ref{FIG:1}. 

Focusing on pentaquark clusters, the right part of figure \ref{FIG:1} shows the rapidity distribution normalised 
to the distribution of baryons, which is relevant for the chances of identifying pentaquarks 
in the debris of a heavy ion collision. At least for SPS energies, this ratio is nearly 
twice as large at $\Delta y = \pm 1$ off midrapidity as at $y = 0$.  For RHIC energies, 
a cylindrical quark gas, thermalised at a temperature of 250~MeV, was assumed as initial condition, 
so that the full dynamics of the expanding fireball is missing. 
The apparently high relative yield of pentaquarks is strongly reduced, of course, by the large size
of the flavour multiplet,
\begin{equation} \label{EQ:3}
\mathbf{3}
\otimes \mathbf{3}
\otimes \mathbf{3}
\otimes \mathbf{3}
\otimes \mathbf{\overline{3}}
 = 
3 \cdot \mathbf{1}
\oplus 8 \cdot \mathbf{8}
\oplus 4 \cdot \mathbf{10}
\oplus 2 \cdot \mathbf{\overline{10}}
\oplus 3 \cdot \mathbf{27}
\oplus \mathbf{35},
\end{equation}
where, assuming SU(3) coupling, only 1 out of 243 states corresponds to the $\Theta^+$, 
while the clustering procedure populates the whole multiplet structure. The left part of figure \ref{FIG:3} 
shows the relative distribution of pentaquark clusters over strangeness. 
The grey band corresponds to the fully SU(3) symmetric case. For SPS energies, there are large deviations, 
corresponding to the strangeness suppression. But even when normalising the relative distribution 
to the actual strangeness content of the quark system (right part of figure \ref{FIG:3}), 
deviations remain for SPS energies. In our model, this is due to the correlations 
carried over into the quark system from the original hadrons. However, we conclude that
the relative strangeness of pentaquarks can give a measure for thermalization and homogenisation 
of the deconfined quark system. 

\begin{figure}[tbhp]
\centerline{\includegraphics[width=0.6\linewidth]{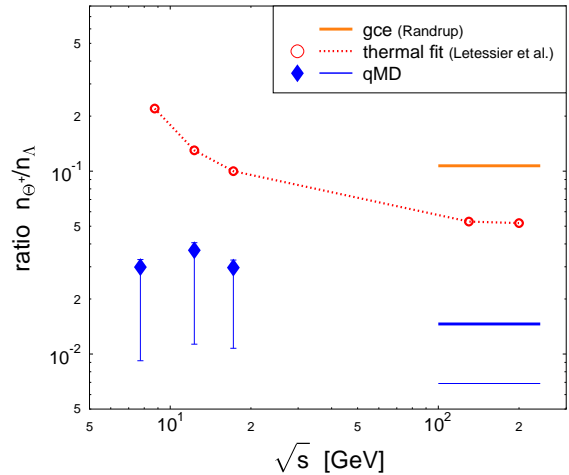}}
\caption{\label{FIG:4}
Ratio $\Theta^+/\Lambda$ as a function of energy from the grand canonical ensemble,
thermal models including strangeness suppression factors, and from qMD clustering.
}
\end{figure}

Finally, we compare the relative yield of $\Theta^+/\Lambda$ as a function of 
bombarding energy from our quark clustering dynamics to the results obtained 
from grand canonical estimates~\cite{Randrup:2003fq} and thermal models including
strangeness suppression factors~\cite{Letessier:2003by} (Figure~\ref{FIG:4}). We find
relative yields roughly one order of magnitude below the earlier estimates. The large 
downward range is due to the use of different coupling schemes, SU(2) for fixed strangeness, 
SU(3), and SU(6) for spin and flavour. It should be noted that the qMD result does not include
hadronic rescattering, which will both destroy $\Theta^+$s formed from quark clustering and
create new ones from hadronic coalescence (\cite{Chen:2003tn} -- there are, however, much 
unknowns about the formation of pentaquarks, see e.\ g.~\cite{Karliner:2004gr}), 
nor the possible population of $\Theta^+$ from decaying $N^*$ pentaquarks (see e.\ g.~\cite{Navarra:2004bj}). 
These factors shift the $\Theta^+/\Lambda$ and require further study.

\ack

The author thanks Marcus Bleicher, Fuming Liu, and Horst St\"ocker for helpful discussions and comments. 
This work was supported by GSI, DFG, and the FIAS.

\end{document}